\documentclass[preprint,11pt]{elsarticle}
\usepackage{dcolumn,amsmath}
\usepackage{graphicx}
\usepackage{multirow} 
\usepackage{wrapfig}
\usepackage{comment}
\usepackage{subcaption}

\newcommand{\wn}{cm$^{-1}$ }

\newcommand{\supd}{$^{d}$}
\newcommand{\supe}{$^{e}$}
\journal{Journal of Molecular Spectroscopy}
\begin{document}
\begin{frontmatter}
\title{Analysis of the $\tilde{A}-\tilde{X}$ bands of the Ethynyl Radical near 1.48$\mu$m and Re-evaluation of $\tilde{X}$ State Energies}

\author[BNL,ASU]{A.~T.~Le}
\ead{Anh.T.Le@asu.edu}
\author[BNL,USB]{Eisen~C.~Gross}
\ead{eisen.gross@stonybrook.edu}
\author[BNL]{Gregory~E.~Hall}
\ead{gehall@bnl.gov}
\author[BNL,USB]{Trevor~J.~Sears \corref{cor1}}
\ead{trevor.sears@stonybrook.edu}
\address[BNL]{Chemistry Division, Brookhaven National Laboratory, Upton, NY 11973-5000, USA}
\address[USB]{Chemistry Department, Stony Brook University, Stony Brook, NY 11794-3400, USA}
\address[ASU]{Now at: School of Molecular Sciences, Arizona State University, Tempe, AZ 85287, USA}
\cortext[cor1]{Corresponding author}

\begin{abstract}
We report the observation and analysis of spectra in part of the near-infrared spectrum of C$_2$H, originating in rotational levels in the ground and lowest two excited bending vibrational levels of the ground $\tilde{X}\,^2\Sigma^+$ state.  In the analysis, we have combined present and previously reported high resolution spectroscopic data for the lower levels involved in the transitions to determine significantly improved molecular constants to describe the fine and hyperfine split rotational levels of the radical in the zero point, $v_2=1$ and the $^2\Sigma^+$ component of $v_2=2$. Two of the upper state vibronic levels involved had not been observed previously.  The data and analysis indicate the electronic wavefunction character changes with bending vibrational excitation in the ground state and provide avenues for future measurements of reactivity of the radical as a function of vibrational excitation. 
\end{abstract}


\end{frontmatter}

\date{\today}


\section{Introduction}
\label{Intro}

The ethynyl radical, C$_2$H,  is an important intermediate in soot formation in hydrocarbon combustion,\cite{ Eiteneer2003,Boullart1996} in the synthesis of single-walled carbon nanotubes,\cite{Omachi2014, Wang2014} and in star-formation regions, late star carbon-containing circumstellar shells, and in the interstellar medium,\cite{Irvine2011} where it often serves as a proxy for the more difficult to detect acetylene. The radical exhibits a variety of reactivity including simple H-abstraction with saturated reaction partners as well as addition followed by collisional stabilization of the internediate, with unsaturated species.\cite{Kovacs2010}    Given its relevance in such a variety of fields, laboratory spectroscopic and kinetics measurements on the radical have been extensive.  A fairly comprehensive list of the relevant spectroscopic work before 2003 can be found in references,\cite{Tarroni2003, Tarroni2004} which describe high level \textit{ab initio} calculations of the vibronic energy levels below 10 000 \wn.  This work has served as a guide to most of the more recent spectroscopic studies which have been referenced in our previous report\cite{Le2016} on the near infrared spectrum of the radical. \\   
Driven by the need for sensitive, non-intrusive, tools for probing the reactivity of the radical, its gas phase infrared (IR) and near-IR spectrum has been extensively studied.  The ground state has $^2\Sigma^+$ symmetry and the Hirota group\cite{Kanamori1988} made a pioneering study of the C$-$C stretching vibration in the mid-IR, while Curl and co-workers\cite{Yan1987,Curl1985,Curl1985erratum} concentrated on the C$-$H stretching region near 3 $\mu$m.  The spectra in this region are far more complicated than can be explained by the vibrations of a $^2\Sigma$ radical and it became clear that the excited $\tilde{A}\,^2\Pi$ electronic state strongly influences the structure of the spectrum.  An overview of the bending vibrational mode structure in the ground state was also obtained in laser-induced fluorescence work by Hsu et al.\cite{Hsu1993,Hsu1995,Chiang1999}\\
Computational studies by Tarroni and Carter\cite{Tarroni2003,Tarroni2004} put the analysis of the experimental data on a more secure footing.  They showed that the electronic wavefunctions for all the excited vibrational levels in the ground $^2\Sigma^+$-state possess some degree of $A\,^2\Pi$-state character. At the energies of the C$-$H stretching vibration in the ground state, the two electronic states are highly mixed, and the $\tilde{A}\,-\,\tilde{X}$ origin band character is shared among at least 6 different bands in the spectrum at these wavelengths. In addition, the  $\tilde{A}\,^2\Pi$-state suffers a Renner-Teller interaction that breaks the $\Pi$-state degeneracy and leads to even more irrgularities in the energy levels.  Subsequent work to shorter, near-IR, wavelengths by the Nesbitt group\cite{Sharp-Williams2011a,Sharp-Williams2011b} remeasured some of the bands previously observed by Curl and coworkers, and assigned others, using a slit-jet source that provided simpler spectra.  Recent work on C$_2$H in our group \cite{Le2016} identified and rotationally assigned three bands originating from the zero point level of the $\tilde{X}\,^2\Sigma$ state to upper states around 6630 and 7135 \wn  by near-IR diode laser transient absorption spectroscopy.  Tarroni and Carter, and matrix spectroscopic measurements by Forney et al.,\cite{Forney1995} showed that bands in this region are the strongest of all bands in the IR and near-IR spectrum of the radical and they will therefore be ideal candidates for future kinetics and dynamics measurements.  They derive their strength from the $\tilde{A}(v_1,v_2,v_3)=(0,0,2)\, -\, \tilde{X}(0,0,0)$ band, which has a large Franck-Condon factor.  The spectra could be rotationally and vibronically assigned using known ground state combination differences\cite{Gottlieb1983} and by comparison to the predictions of Tarroni and Carter.\cite{Tarroni2004} All the spectral features assigned in our earlier work are due to absorption from the ground state, but many additional bands were clearly present in the observed spectra, particularly in samples recorded at times before collisinal relaxation simplified the spectrum. In the present paper, we report ro-vibronic assignments in the same spectral region for transitions originating in vibrationally hot bending levels and have also extended the wavelength coverage to record a further $^2\Pi - ^2\Sigma$ band, near 6820 \wn, originating in the ground state.  \\

\section{Experiment}
Details of the spectrometer used here have been given previously.\cite{Le2016} A Sacher Lasertecknik tunable diode laser, model TEC 500, with output centered at 1450 nm was used to record Doppler-limited spectra between 6600 and 6900 \wn, following 193 nm ArF excimer laser photolysis of 3,3,3 trifluoropropene, CF$_3$C$_2$H, as a source of C$_2$H. For most of the work, a dual beam setup was employed to reduce laser source noise and the signal and reference beams were imaged on to the elements of an \textit{a.c.}-coupled dual InGaAs photoreceiver (New Focus 1617). A spectrum was obtained by averaging 40 (typically) photolysis laser shots per wavelength step of, normally, 0.005 \wn  in a 1:1 mixture of CF$_3$C$_{2}$H and argon at a total pressure of 1.0 Torr with a detection time gate of 1.0 $\mu$s starting at either zero time delay following the photolysis or after few $\mu$s of delay time to allow for partial collisional relaxation of the initially hot sample. The newly observed $\tilde{A}\,^2\Pi - \tilde{X}\,^2\Sigma(0,0,0)$ band near 6820 \wn was recorded using a frequency-modulated laser beam and heterodyne detection, similar to our previous work on CH$_2$ near-IR spectra.\cite{Chang2011} The chemical conditions were also changed, with a 10$\times$ lower overall precursor  concentration and lower overall pressure, see below.  These changes resulted in improved signal-to-noise ratios compared to the dual beam setup used previously.\cite{Le2016}  

\section{Results and Analysis}
\subsection{Results}

\begin{figure}[t]
	\centering
	\includegraphics[width=1\textwidth]{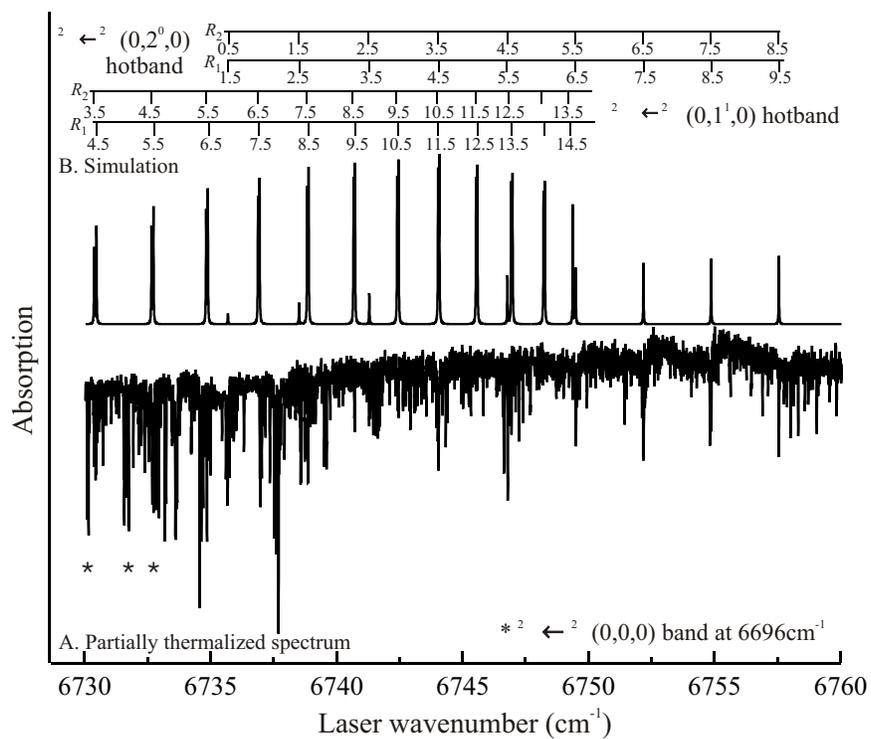}
	\caption{A: A 30 \wn  section of the thermally relaxed observed spectrum at 2.6 $\mu$s time delay following the photolysis laser pulse in a 50\% mixture of precursor in argon at 1 Torr total pressure. B: Simulated spectra including the $^2\Sigma^+$(7088 \wn) $\leftarrow \, \tilde{X}(0,1^1,0)\,^2\Pi$ and $^2\Sigma^+$ (7527 \wn)$ \leftarrow \, \tilde{X}(0,2^0,0)\, ^2\Sigma^+$  transitions were obtained using the optimized set of parameters in Tables \ref{table1} and \ref{table2}.  Lines marked with an asterisk show the previously assigned $\tilde{X}(0,8^{0},2)\,^2\Sigma^+ \leftarrow \, \tilde{X}(0,0,0)\,^2\Sigma^+$ transitions. The upper markers show the assignments for the two hot bands. }
	\label{figure1.}
\end{figure}

\begin{figure}[t]
	\centering
	\includegraphics[width=1\textwidth]{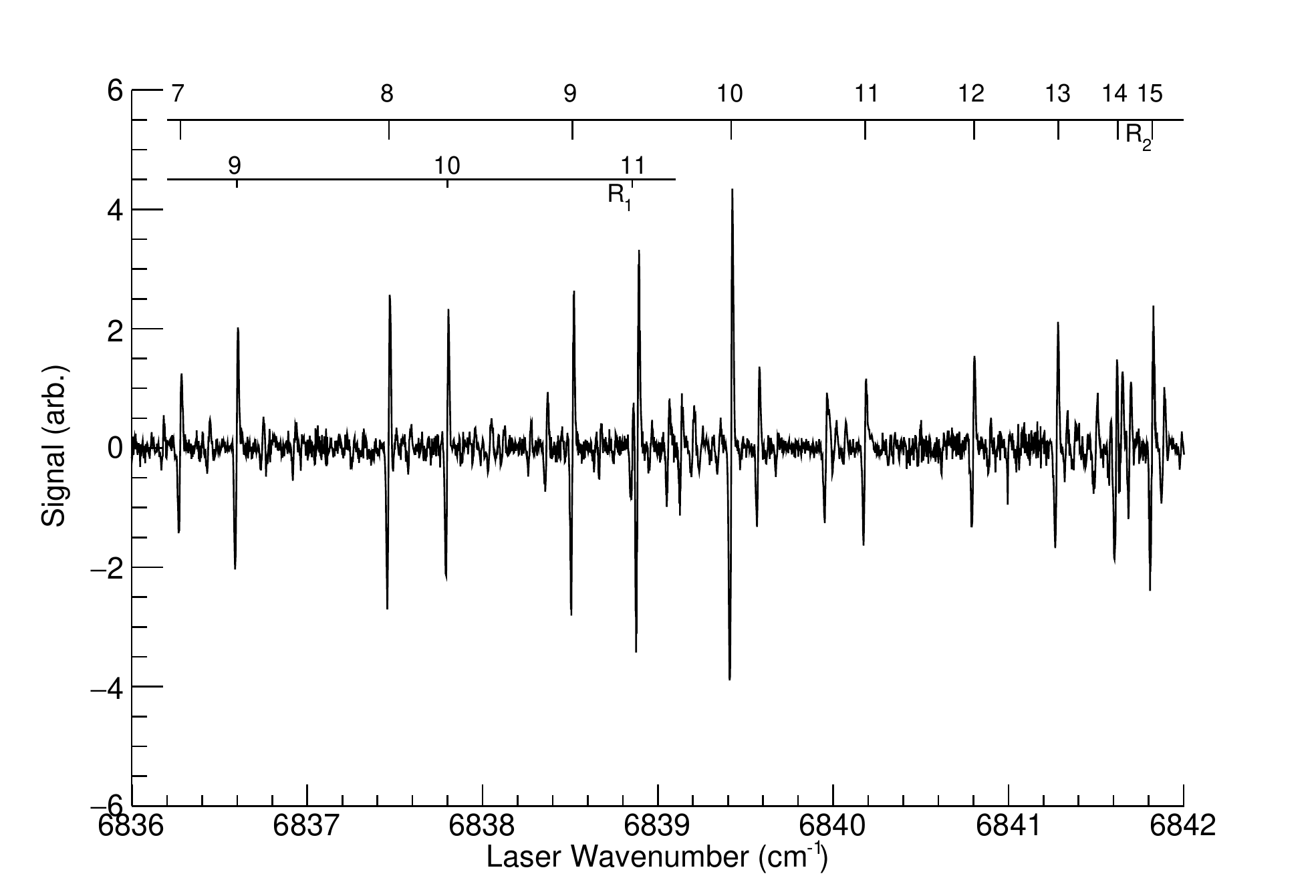}
	\caption{Section of the $^2\Pi$(6819.3) $\leftarrow \tilde{X}(0,0,0)\,^2\Sigma^+$ transition, showing some rotational assignments in a thermally relaxed sample. The experimental conditions were 500 mTorr total pressure of a 10\% mixture of precursor in argon, and plotted is the spectrum at 1 $\mu$sec delay and 1 $\mu$sec detection gate. Lineshapes are first derivatives of the absorption. The R$_1$ series suffers a strong perturbation above R$_1$(11), while the R$_2$ lines are regular and form a band head near N=15.}
	\label{c2h6819-2}
\end{figure}

Spectra in the region covered by the present data were partially assigned previously\cite{Le2016} but searches for hot band transitions were unsuccessful at the time because they used combination differences based on the rotational and fine structure parameters for the lower level reported by Kanamori et al.\cite{Kanamori1988} The sign of $\Lambda$-doubling parameters $p$, $q$, and $q_{D}$ reported by Kanamori et al., and also in work on different bands by Curl et al.\cite{Curl1985,Curl1985erratum} assumed a convention appropriate for vibrational $\ell$-doubling resulting in a reversal in sign compared to the case(\textit{a}) electronic Hamiltonian used here.  This confusion was first pointed out by Hsu and co- workers.\cite{Hsu1993,Hsu1995} The sign convention used by Hsu et al., and ourselves, is also consistent with the older millimeter wave\cite{Woodward1987} and far-infrared laser magnetic resonance\cite{Brown1988} measurements, but the sign confusion was not recognized until much later.\\
With the revised $\Lambda-$doubling sign, a re-examination of the earlier spectra has permitted many new assignments to be made.  As an example, a 30 \wn  section of the observed spectra shown at 2.6 $\mu$s time delay between photolysis and detection to allow for partial rotational relaxation, is shown in Figure \ref{figure1.}.  Lines marked with an asterisk were previously assigned to \textit{R}$ _{1} $ and \textit{R}$ _{2}$  branches to a $^2\Sigma^+ $ state at 7088 cm$^{-1}$ from the $\tilde{X}(0,0,0)\, ^2\Sigma ^+$ ground state.\cite{Le2016}  An additional  $^2\Sigma^+ \leftarrow\, ^2\Sigma ^+$ band pattern was recognized, and assigned to transitions originating from the $v_2^{\prime\prime}=2$, $\tilde{X}(0,2^{0},0)\,^2\Sigma^+$ state.   The \textit{R$_{1}$} and \textit{R$_{2}$} branches of the $^2\Sigma^+ \leftarrow \,\tilde{X}(0,1^{1},0)\,^2\Pi $ and the $^2\Sigma^+ \leftarrow \,\tilde{X}(0,2^{0},0)\,^2\Sigma^+$ band are both indicated on the figure. Details of all the transitions assigned in the present work are available as supplementary data for this paper.\cite{Supplemental} \\
A combination differences search program was written assuming parameters from Hsu et al.\cite{Hsu1993} to identify hot band transitions originating from the $(0,1^1,0)\,^2\Pi$ symmetry level to either a $^2\Pi$- or $^2\Sigma^+$-symmetry state. With a small negative value for the $\tilde{X} (0,1^1,0)$ spin-orbit constant, $A$,\cite{Woodward1987} the case (\textit{a}) $^2\Pi_{3/2}$ and $^2\Pi_{1/2}$ energy levels map onto F$_1$ and F$_2$, respectively, in a case(\textit{b}) representation.\cite{HerzbergDiat} Six main branches (\textit{P$_1$}, \textit{P$_2$}, \textit{Q$_1$}, \textit{Q$_2$}, \textit{R$_1$}, and \textit{R$_2$}) are expected for $^2\Sigma^+ \leftarrow \,^2\Pi $ transitions, similar to the $^2\Pi \leftarrow \,^2\Sigma^+ $ transitions previously assigned. Four weaker satellite branches, with F$_1$ $\leftrightarrow$ F$_2$ were not observed.  Fortrat diagrams \cite{HerzbergDiat} for the expected bands based on the known lower level rotational combination differences were found to be useful aids in making assignments.\\
Spectra in the region between 6780 \wn and 6870 \wn were recorded more recently and typical results are shown in Figure \ref{c2h6819-2}.  The improved signal-to-noise ratio compared to Fig. \ref{figure1.} is due to a combination of the change to FM detection and, mainly, more favorable chemical conditions at the lower pressure and concentrations used, as detailed in the figure captions.

\subsection{Analysis}
\subsubsection{The $^2\Sigma^+ (7088) \leftarrow \tilde{X}(0,1^1,0)\,^2\Pi$ band}
\label{subsub1}
 In total, 77 transitions from rotational levels in the $\tilde{X}(0,1^1,0)\,^2\Pi$ state to those in the previously observed\cite{Le2016} $^2\Sigma $ state at 7088 \wn  were identified via lower state combinational differences, initially using the constants reported by Hsu et al.\cite{Hsu1993} The upper state in this transition suffers strong perturbations that were identified in the earlier work.\cite{Le2016}   We have combined all the published high precision spectroscopic data for the lower $\tilde{X}(0,1^1,0)\,^2\Pi$ level in a weighted least squares fit to the lower state constants.  We included the microwave rotational transition frequencies, with their hyperfine splittings, reported by Woodward et al.\cite{Woodward1987} and combination differences extracted from the mid-infrared data of Kanamori and Hirota\cite{Kanamori1988} that originate in the same lower level, as well as an extensive set of lower state combination differences from the present measurements.  The data were weighted as the inverse square of the measurement uncertainties.  Woodward's \cite{Woodward1987} data explicitly lists estimated measurement errors for each point, while the infrared combination differences\cite{Kanamori1988} were assumed to have an uncertainty of $\pm0.001$ \wn and the present ones $\pm0.002$ \wn.\\
 The $^2\Pi$ Hamiltonian used was that of Kawaguchi et al.,\cite{Kawaguchi1985} but including a spin-rotation coupling term $H_{sr}=\gamma\mathbf{J.S}$ more appropriate for the case(\textit{b}) nature of the state, instead of terms representing centrifugal distortion corrections to the spin-orbit coupling.  Brown and Watson\cite{Brown1977} showed that energy contributions from these two types of terms are not distinguishable.  Higher order corrections to the lambda-doubling terms were included using the matrix procedure on page 546 of Brown and Carrington.\cite{BandC2003}  The resulting molecular constants for the $\tilde{X}\,^2\Pi\,(0,1^1,0)$ level are listed in Table \ref{table1}.  The centrifugal distortion correction to the $\Lambda-$doubling parameter \textit{p} was not determined and fixed at zero.  A non-zero value for this parameter was reported by Woodward \textit{et al.},\cite{Woodward1987} but with a $1\sigma$ uncertainty twice as large as the parameter value.  The other parameters are very close to those reported by Woodward et al., but are slightly better determined.  Most importantly, they reproduce all the observed $\tilde{X}\,^2\Pi\,(0,1^1,0)$ levels to within the measurement uncertainties.\\
For the upper $^2\Sigma^+ (7088)$ level in this transition, the energies and molecular parameters were determined in our earlier work\cite{Le2016} via transitions from the $\tilde{X}^2\Sigma^+$ zero point level. In the course of the present analyses, we have refit these data using a refined set of molecular constants for the zero point level obtained in a combined, weighted, fit of the known microwave and astronomical data\cite{Gottlieb1983} and combination differences from the near-infrared data that, although less precise, extend to much higher rotational levels.  The resulting parameters are given in the first column of Table \ref{table1} and reproduce all the rotational levels up to N=22 to within their measurement precision.  With these lower state constants, data for transitions to unperturbed upper state rotational levels in the upper state\cite{Le2016} were refit yielding the parameters in Table \ref{table2}.  Combining these results, transition frequencies for all unperturbed assigned lines in the present band, were well reproduced with a shift of +0.003 \wn in the upper state energy compared to the value determined from the data for the band from the $\tilde{X}^2\Sigma^+\, (000)$ level.  The shift could also be accommodated by a similar (negative) change in the assumed\cite{Kanamori1988} $\tilde{X}(0,1^1,0)\,^2\Pi\,$ energy of 371.6034 \wn.  This wavenumber difference is marginally greater than the estimated measurement error, and must be due to an absolute wavenumber calibration inaccuracy between the various measurements.  The supplementary information for this paper\cite{Supplemental} also includes calculated rotational energy levels for all the vibronic levels involved in the observed transitions based on the spectroscopic constants in Tables \ref{table1} and \ref{table2}. 

\subsubsection{The $^2\Sigma^+(7527) \leftarrow \tilde{X}(0,2^0,0)\,^2\Sigma^+$ band}
\label{subsub2}
In a similar way, 44 measured transition wavenumbers for the $^2\Sigma^+ $ (7527 cm$^{-1})$ $ \leftarrow\, \tilde{X}(0,2^0,0)\,^2\Sigma ^+$ band were identified via combination differences using the known\cite{Killian2007} $\tilde{X}(0,2^0,0)\,^2\Sigma ^+ $ state parameters. During the analysis, it became clear that higher spin-rotational intervals in the $\tilde{X}\,^2\Sigma ^+\,(0,2^0,0)$ lower state were systematically different to those computed using the reported \cite{Killian2007} molecular constants.  Similar to what was done for the band above, ground state combination differences from the present measurements were combined with the more precisely measured rotational intervals for the lowest few rotational levels of the lower $\tilde{X}(0,2^0,0)\,^2\Sigma ^+$ state in a weighted least squares fit to determine the best fit lower state molecular parameters.  This process brought to light the fact that the lowest frequency rotational transition reported in this vibrational level was inconsistent with the remainder of the data, being calculated 1.14 MHz above the published frequency using the best fit constants in Table \ref{table1}.  Killian et al.\cite{Killian2007} reported only one hyperfine component in the N $= 1\leftarrow0$ rotational transition of $\tilde{X}(0,2^0,0)$ C$_2$H.  This rotational transtion is expected (and seen, in other vibrational levels) to possess five observable hyperfine components.  On the basis of the rest of the available data, the reported line frequency does not match any possible hyperfine component of this transition in $\tilde{X}^2\Sigma ^+ (0,2^0,0)$.  In the original report,\cite{Killian2007} only N $= 1\leftarrow0\, , 2\leftarrow1\, \mathrm{and} \, 3\leftarrow2\,$ transtions were observed in this vibrational level, and the discrepancy in the lowest one was hidden because 4 rotational and spin-rotational constants were varied to fit the observed energy intervals. The original fitting resulted in an unphysical negative value for the centrifugal distotion constant, \textit{D}, and an additional term representing the centrifugal distortion correction the the spin-rotation coupling, $\gamma_D$, that was not needed for any of the other observed vibrational levels. \\
Finally, the microwave data, excluding the N=$1 \leftarrow 0$ transition, were combined with 43 near infrared transitions terminating in upper state levels determined to be unperturbed in a weighted least squares fit to determine upper and lower state molecular constants. The results are summarized in the two tables. The perturbed transitions are noted in the appropriate tables in the supplementary data for this paper.\cite{Supplemental} The upper state is 6733 \wn  above the $\tilde{X}\,^2\Sigma ^+\,(0,2^0,0) $ state.  The upper state centrifugal distortion constant is negative, but this may plausibly be the case in the highly mixed and perturbed vibronic levels at this energy in C$_2$H.  Overall, this level shows much smaller levels of perturbation than the lower energy $^2\Sigma^+\,(7088)$ state described in section \ref{subsub1}. The lower level rotational structure is more regular than implied by the results in reference,\cite{Killian2007} although one higher centrifugal distortion constant ($H$) was required to satisfactorily model the measurements compared to the two lower vibrational levels in table \ref{table1}.

\subsubsection{The $^2\Pi(6819) \leftarrow \tilde{X}(0,0,0)\,^2\Sigma^+$ band}
\label{subsub3}
Spectra near 1.47 $\mu$m showed several clear progressions that have been assigned to a new band corresponding to absorption from the ground state of the radical and terminating in a previously unobserved $^2\Pi$ state.  Combination differences permitted unambigous assignments to the main P- and R-branch series from the two fine structure components of the ground $^2\Sigma$ state.  Figure \ref{c2h6819-2} shows part of the R-branch series including the band head region. From the figure, it is clear that the R$_1-$branch suffers a severe perturbation around R$_1$(12) where the regular progression observed from lower rotational quantum numbers ceases.  The spectral coverage did not extend to low enough frequencies to permit the observation of the corresponding P$_1-$branch lines.  In fact, fitting of the data showed that the R$_1$(8), R$_1$(9) and R$_1$(10) lines were progressively shifted away from their predicted positions in the spectrum.  By contrast, the R$_2-$branch lines are quite regular.  There is a well-known problem making unanbiguous assignments to the Q$-$branch lines in this band type, because they terminate in the other $\Lambda-$doublet component to the P and R-lines.  Assignments were made by assuming the smallest magnitude $\Lambda-$doubling at the lowest rotational quantum number.  Changing the spectral assignments by $\pm1$ in N resulted in much poorer fits.\\
In total, 72 rotational transitions have been assigned and the data are available as supplementary data.\cite{Supplemental} Since the lower state levels are now very well known, a least squares fit to spectroscopic constants for the upper state energy levels was performed. The results are summarized in the third column of Table \ref{table2}.  The position of the origin corresponds closely to a level of $^2\Pi$ symmetry calculated\cite{Tarroni2003} at 6824.8 \wn and identified as a level with a dominant contributions of $\tilde{X}(2,1^1,0),\tilde{A}(0,0,2)^1$ to the electronic wavefunction.  The fitting required a number of higher order distortion-type constants but, even so, the levels were not fit to the expected measurement precision with a fit variance of 4.4 times the expected based on measurement uncertainties.  Small local and global perturbations present in the rotational energy level structure are the reason for this. 
 
\section{Discussion and Conclusions}

In this and earlier\cite{Le2016}  work, we have analyzed data covering a small region of the extensive\cite{Tarroni2003, Tarroni2004} near-infrared spectrum of the C$_2$H radical. The assigned transitions mostly originate in bending vibrationally hot levels of the ground state and terminate in a $^2\Sigma^+$ state lying at 7088 \wn above the zero point $\tilde{X}\,^2\Sigma^+$ level and a second $^2\Sigma^+$ level at 7527 \wn.  Both upper levels are predicted to be strongly vibronically mixed.\cite{Tarroni2004}  The first has previously been spectroscopically observed\cite{Le2016} but the higher energy one has not previously been reported.  Tarroni and Carter\cite{Tarroni2004} computed a level of the correct symmetry approximately 20 \wn above the observed band origin and identified the dominant contribution to the wavefunction to be $\tilde{X}(1,2^0,2)\,^2\Sigma^+$.  \\
The lower levels in the assigned transitions have been spectroscopically characterized previously, but we have combined data in the literature with the present measurements to refine the spectroscopic parameters describing the fine and hyperfine split rotational levels so that the new molecular constants reproduce all the known levels up to above N = 20 to within their measured precision.  These results permit reliable modelling of spectra involving these levels at temperatures up to and slightly above ambient.  In the process, a misassignment of a microwave transition in the $\tilde{X}(0,2^0,0)\,^2\Sigma^+$ vibrational level was identified.  Removing this one line resulted in a correction of the centrifugal distortion constant that had previously been reported as negative, which prevented accurate calculation of higher rotational level energies in this vibrational level.  Because the $^2\Sigma^+\,(7088)$ upper state in this work was also connected to the zero point level of the ground state by our earlier work, we have similarly refined and improved the ground state molecular parameters for the radical.  \\
In open-shell linear molecules such as CCH, the vibrational and the electronic orbital angular momenta are strongly coupled by the Renner-Teller effect, and $\Lambda$-type and $\ell$-type doubling combine to produce a resultant $K$-type doubling.\cite{BrownRTE}  $\Lambda$-type doubling in a  $^2\Pi$ electronic state is described by two parameters \textit{p} and \textit{q}, but $\ell$-type doubling requires only one, \textit{q}.  In an isolated $^2\Pi$ electronic state, the $\Lambda$-doubling constant \textit{p} is usually much greater than \textit{q} as it is in the case of the mixed  $^2\Pi$ upper states (Table \ref{table2}). However, \textit{p$\,\ll$q} in the $\tilde{X}(0,1^{1},0)\,^2\Pi$ state indicating that the vibrational contribution to \textit{q} is much larger than the electronic contribution in this case, not entirely surprising because this level lies deep in the $\tilde{X}$ state potential and has been calculated to possess 95\% $\tilde{X}\,^2\Sigma^+$ character.\cite{Tarroni2003} \\
The variation of the hyperfine parameter $b$ with vibration in the $\tilde{X}$ state is interesting. Since the spin-spin dipolar contribution to the hyperfine coupling ($c$) does not change significantly, the variation must be arising due to changes in the Fermi-contact parameter, $b_F$, which is\cite{BandC2003} $b_F=b+c/3$.  In low-lying vibrational levels such as these, one does not normally expect much variation in hyperfine coupling with vibrational excitation.  The marked reduction in size of the Fermi-contact contribution to the hyperfine coupling here imples a decrease in sigma character of the electronic wavefunction within 700 \wn of excitation above the zero point level.  However, Tarroni and Carter\cite{Tarroni2003} find the $\Pi$-state character of the wavefunction does not change between the $\tilde{X}(0,1^1,0)\,^2\Pi$ and $\tilde{X}(0,2^0,0)\,^2\Sigma ^+$ levels: both are calculated to possess about 5\% $\Pi$ character, so this apparently cannot explain the observed variation in the size of the Fermi-contact parameter. \\ 
In summary, we have reported new assignments in the near infrared spectrum of this radical and improved the precision of all the spectroscopic constants for the zero-point and two low-lying vibrational levels in the ground electronic state.  Aside from their intrinsic spectroscopic interest, the new data will be useful for future kinetic and dynamical measurements in the many reaction systems where C$_2$H plays a role.

\section*{Acknowledgements}

Work at Brookhaven National Laboratory was carried out under Contract No. DE-SC0012704 with the U.S. Department of Energy, Office of Science, and supported by its Division of Chemical Sciences, Geosciences and Biosciences within the Office of Basic Energy Sciences. 

\small
\bibliography{HotCCH.bib}

\begin{thebibliography}{10}
\expandafter\ifx\csname url\endcsname\relax
  \def\url#1{\texttt{#1}}\fi
\expandafter\ifx\csname urlprefix\endcsname\relax\def\urlprefix{URL }\fi
\expandafter\ifx\csname href\endcsname\relax
  \def\href#1#2{#2} \def\path#1{#1}\fi

\bibitem{Eiteneer2003}
B.~Eiteneer, M.~Frenklach, Experimental and modeling study of shock-tube
  oxidation of acetylene, Int. J. Chem. Kin. 35~(9) (2003) 391--414.
\newblock \href {http://dx.doi.org/10.1002/kin.10141}
  {\path{doi:10.1002/kin.10141}}.

\bibitem{Boullart1996}
W.~Boullart, K.~Devriendt, R.~Borms, J.~Peeters, Identification of the sequence
  {CH($^2\Pi$)} + {C$_2$H$_2$ $\rightarrow$ C$_3$H$_2$ + H (and C$_3$H +
  H$_2$)} followed by {C$_3$H$_2$ + O $\rightarrow$ C$_2$H + HCO (or H+CO)} as
  {C$_2$H source in C$_2$H$_2$/O/H} atomic flames, J. Phys. Chem. 100 (1996)
  998--1007.

\bibitem{Omachi2014}
H.~Omachi, T.~Nakayama, E.~Takahashi, Y.~Segawa, K.~Itami, Initiation of carbon
  nanotube growth by well-defined carbon nanorings, Nature Chem. 5~(7) (2013)
  572--576.
\newblock \href {http://dx.doi.org/10.1038/NCHEM.1655}
  {\path{doi:10.1038/NCHEM.1655}}.

\bibitem{Wang2014}
Y.~Wang, X.~Gao, H.-J. Qian, Y.~Ohta, X.~Wu, G.~Eres, K.~Morokuma, S.~Irle,
  Quantum chemical simulations reveal acetylene-based growth mechanisms in the
  chemical vapor deposition synthesis of carbon nanotubes, Carbon 72 (2014)
  22--37.
\newblock \href {http://dx.doi.org/10.1016/j.carbon.2014.01.020}
  {\path{doi:10.1016/j.carbon.2014.01.020}}.

\bibitem{Irvine2011}
W.~M. Irvine, Ethynyl radical, in: M.~Gargaud, R.~Amils, J.~C. Quintanilla,
  H.~J.~J. Cleaves, W.~M. Irvine, D.~L. Pinti, M.~Viso (Eds.), Encyclopedia of
  Astrobiology, Springer Berlin Heidelberg, Berlin, Heidelberg, 2011, p. 507.

\bibitem{Kovacs2010}
T.~Kov\'acs, M.~A. Blitz, P.~W. Seakins, H-atom yields from the photolysis of
  acetylene and from the reaction of {C}$_2${H} with {H}$_2$, {C}$_2${H}$_2$,
  and {C}$_2${H}$_4$, J. Phys. Chem. A 114 (2010) 4735--4741.

\bibitem{Tarroni2003}
R.~Tarroni, S.~Carter, Theoretical calculation of vibronic levels of {C$_2$H}
  and {C$_2$D} to 10 000 cm$^{-1}$, J. Chem. Phys 119~(24) (2003) 12878--12889.

\bibitem{Tarroni2004}
R.~Tarroni, S.~Carter, Theoretical calculation of absorption intensities of
  {C$_2$H} and {C$_2$D}, Molec. Phys. 102~(21-22) (2004) 2167--2179.

\bibitem{Le2016}
A.~T. Le, G.~E. Hall, T.~J. Sears, The near-infrared spectrum of ethynyl
  radical, J. Chem. Phys. 145~(7).
\newblock \href {http://dx.doi.org/10.1063/1.4961019}
  {\path{doi:10.1063/1.4961019}}.

\bibitem{Kanamori1988}
H.~Kanamori, E.~Hirota, Vibronic bands of the {CCH} radical observed by
  infrared diode-laser kinetic spectroscopy, J. Chem. Phys. 89~(7) (1988)
  3962--3969.
\newblock \href {http://dx.doi.org/10.1063/1.454877}
  {\path{doi:10.1063/1.454877}}.

\bibitem{Yan1987}
W.-B. Yan, C.~Dane, D.~Zeitz, J.~L. Hall, R.~Curl, Color center laser
  spectroscopy of {C$_2$H} and {C$_2$D}, J. Molec. Spectrosc. 123~(2) (1987)
  486 -- 495.
\newblock \href {http://dx.doi.org/10.1016/0022-2852(87)90294-3}
  {\path{doi:10.1016/0022-2852(87)90294-3}}.

\bibitem{Curl1985}
R.~F. Curl, P.~G. Carrick, A.~J. Merer, Rotational analysis of the
  $\tilde{A} \leftarrow \tilde{X}$ system of {C$_2$H}, J. Chem. Phys.
  82~(8) (1985) 3479--3486.
\newblock \href {http://dx.doi.org/10.1063/1.448927}
  {\path{doi:10.1063/1.448927}}.

\bibitem{Curl1985erratum}
R.~F. Curl, P.~G. Carrick, A.~J. Merer, Erratum: Rotational analysis of the
  $\tilde{A} \leftarrow \tilde{X}$ system of {C$_2$H} [{J}. {C}hem. {P}hys.
  82, 3479 (1985)], J. Chem. Phys 83~(8) (1985) 4278--4278.
\newblock \href {http://dx.doi.org/10.1063/1.449873}
  {\path{doi:10.1063/1.449873}}.

\bibitem{Hsu1993}
Y.-C. Hsu, J.~J.-M. Lin, D.~Papousek, J.-J. Tsai, The low-lying bending
  vibrational levels of the {CCH (X)} radical studied by laser-induced
  fluorescence, J. Chem. Phys. 98 (1993) 6690--6696.

\bibitem{Hsu1995}
Y.-C. Hsu, Y.-J. Shiu, C.-M. Lin, Laser induced fluorescence spectroscopy of
  {C$_2$H ($X^2\Sigma^+$)} in vibrationally excited levels up to 4500
  cm$^{-1}$, J. Chem. Phys 103~(14) (1995) 5919--5930.

\bibitem{Chiang1999}
W.-Y. Chiang, Y.-C. Hsu, Laser spectroscopy of {CCH} in the $36600\, {-}
  \,39700$ cm$^{-1}$ region, J. Chem. Phys. 111 (1999) 1454--1461.

\bibitem{Sharp-Williams2011a}
E.~N. Sharp-Williams, M.~A. Roberts, D.~J. Nesbitt, Dark state vibronic
  coupling in the {$\tilde{A}\,^2\Pi \leftarrow \tilde{X}\,^2\Sigma^+$} band of
  ethynyl radical via high resolution infrared absorption spectroscopy, Phys.
  Chem. Chem. Phys. 13~(39) (2011) 17474--17483.
\newblock \href {http://dx.doi.org/10.1039/c1cp21523j}
  {\path{doi:10.1039/c1cp21523j}}.

\bibitem{Sharp-Williams2011b}
E.~N. Sharp-Williams, M.~A. Roberts, D.~J. Nesbitt, High resolution slit-jet
  infrared spectroscopy of ethynyl radical: {$^2\Pi\,-\, ^2\Sigma^+$} vibronic
  bands with sub-{D}oppler resolution, J. Chem. Phys. 134~(6) (2011) 064314.
\newblock \href {http://dx.doi.org/10.1063/1.3532088}
  {\path{doi:10.1063/1.3532088}}.

\bibitem{Forney1995}
D.~Forney, M.~E. Jacox, W.~E. Thomson, The infrared and near infrared spectra
  of {HCC} and {DCC} trapped in solid neon, J. Molec. Spectrosc. 170 (1995)
  178--214.

\bibitem{Gottlieb1983}
C.~A. Gottlieb, E.~W. Gottlieb, P.~Thaddeus, Laboratory and astronomical
  measurement of the millimeter wave spectrum of the ethynyl radical {CCH},
  Astrophys. J. 264~(2) (1983) 740--745.
\newblock \href {http://dx.doi.org/10.1086/160647} {\path{doi:10.1086/160647}}.

\bibitem{Chang2011}
C.-H. Chang, Z.~Wang, G.~E. Hall, T.~J. Sears, J.~Xin, Transient laser
  absorption spectroscopy of {CH}$_2$ near 780nm, J. Molec. Spectrosc. 267~(1)
  (2011) 50 -- 57.
\newblock \href {http://dx.doi.org/https://doi.org/10.1016/j.jms.2011.02.004}
  {\path{doi:https://doi.org/10.1016/j.jms.2011.02.004}}.

\bibitem{Woodward1987}
D.~R. Woodward, J.~C. Pearson, C.~A. Gottlieb, M.~Gu\'elin, P.~Thaddeus,
  Laboratory study of the rotational spectrum of vibrationally excited
  {C$_2$H}, Astron. Astrophys. 186 (1987) L14--L18.

\bibitem{Brown1988}
J.~M. Brown, K.~M. Evenson, The far-infrared laser magnetic resonance spectrum
  of vibrationally excited {C$_2$H}, J. Molec. Spectrosc. 131~(1) (1988) 161 --
  171.
\newblock \href {http://dx.doi.org/10.1016/0022-2852(88)90115-4}
  {\path{doi:10.1016/0022-2852(88)90115-4}}.

\bibitem{Supplemental}
A.~T. Le, E.~C. Gross, G.~E. Hall, T.~J. Sears, Supplementary data for this
  paper including tables of assigned transitions and computed energy levels is
  available from the {J}ournal.

\bibitem{HerzbergDiat}
G.~Herzberg, Molecular Spectra and Molecular Structure: {I}. Spectra of
  Diatomic Molecules, 2nd Edition, Van Nostrand Reinhold Company, Inc., 1950.

\bibitem{Kawaguchi1985}
K.~Kawaguchi, S.~Saito, E.~Hirota, Microwave spectroscopy of the {NCO} radical
  in the \textit{v}$_2$=0 $^2{\Pi}$, \textit{v}$_2$=1 $^2{\Delta}$,
  \textit{v}$_2$=2 $^2{\Phi}$ vibronic states, Molec. Phys 55~(2) (1985)
  341--350.

\bibitem{Brown1977}
J.~M. Brown, J.~K.~G. Watson, Spin-orbit and spin-rotation coupling in doublet
  states of diatomic molecules, J. Molec. Spectrosc. 65~(1) (1977) 65 -- 74.

\bibitem{BandC2003}
J.~M. Brown, A.~Carrington, Rotational Spectroscopy of Diatomic Molecules, 2nd
  Edition, Cambridge University Press, Inc., Cambridge, UK, 2003.

\bibitem{Killian2007}
T.~C. Killian, C.~A. Gottlieb, P.~Thaddeus, Rotational spectra of vibrationally
  excited {CCH} and {CCD}, J. Chem. Phys 127~(11) (2007) 114320.
\newblock \href {http://dx.doi.org/10.1063/1.2768927}
  {\path{doi:10.1063/1.2768927}}.

\bibitem{BrownRTE}
J.~M. Brown, F.~Jorgensen, Vibronic energy levels of a linear triatomic
  molecule in a degenerate electronic state: a unified treatment of the
  {R}enner-{T}eller effect, in: I.~Prigogine, S.~A. Rice (Eds.), Adv.Chem.
  Phys., Vol.~52, Wiley, New York, 1983, pp. 117--179.
\newblock \href {http://dx.doi.org/10.1002/9780470142769.ch2}
  {\path{doi:10.1002/9780470142769.ch2}}.

\end{thebibliography}

\clearpage

\begin{table}[t]
		\centering
		\caption{Spectroscopic parameters in MHz for the lower states in the observed C$_2$H bands$^a$}
	\label{table1}
	        \small
{\begin{tabular}{llll}
		\hline
		\hline
		\  & $\tilde{X}(0,0,0)$  & $\tilde{X}(0,1^1,0)$ & $\tilde{X}(0,2^0,0)$  \\
		\   &$^{2}\Sigma ^{+}$ &  $^{2}\Pi  $ & $^{2}\Sigma ^{+}$ 	\\
		\hline
		$A$                          &                & 10391.15(24)   &                   \\
		$B$                          & 43674.5370(3) & 43508.4504(73) & 43548.314(20)     \\
		$D$                          & 0.106851(35)    & 0.11459(34)     & 0.0949(14)       \\
		$10^2\,H $              &                 &                 & 0.8162(11)       \\
		$ \gamma $                   & -62.5947(12)     & -46.735(60)   & -34.75(16)      \\
		$10^2\, \gamma_{D}$    & -0.496(37)      &               &  -1.93(80)      \\
		$p$                          &                & -4.354(38)    &                  \\
		$q$                          &                & -340.5142(96) &                     \\
		$10^2\,q_D$               &                  &  0.208(60)     &                   \\
		$a-(b+c)/2 ^b$             &                  &  -21.698(76)   &                   \\
		$a+(b+c)/2 ^b$                &                  &  22.63(12)     &                   \\
		$b$                        &  40.4255(38)      &  31.08(10)     &  26.61(17)       \\
		$c$                        &  12.2488(48)       &                &  12.76(28)       \\
		$d$                        &                  &  3.490(67)     &                   \\
		Origin$^{c}$               & 0                &  371.6034(3)$^d$  & 794.300(6)$^e$   \\
		\hline
		&\multicolumn{2}{c}{Calculated values$ ^{f} $}  \\
		\hline
		Origin$^{c}$               & 0             & 371.4         & 794.3               \\		
		\hline
		\hline
		\multicolumn{4}{l}{\textit{a.} The numbers in parenthesis are one standard }\\
		\multicolumn{4}{l}{deviation of the fit in units of the last quoted decimal place.}\\
         \multicolumn{4}{l}{\textit{b.} Combination of hyperfine parameters.  } \\
         \multicolumn{4}{l}{ Derived constants: $a=0.47(14)$, $c=13.25(17)$ MHz.} \\
	\multicolumn{4}{l}{\textit{c.} In wavenumbers (\wn) }\\
        \multicolumn{4}{l}{\textit{d.} From Reference \cite{Kanamori1988}}\\
        \multicolumn{4}{l}{\textit{e.} From Reference \cite{Chiang1999}} \\
        \multicolumn{4}{l}{\textit{f.} Predicted values are from Tarroni and Carter \cite{Tarroni2003}} \\	
	\end{tabular}}
	\normalsize
\end{table}

\newpage

\begin{table}[t]
		\centering
		\caption{Spectroscopic parameters in wavenumbers (\wn ) for the upper levels in the observed near-IR C$_2$H bands$^{a,b}$}
	\label{table2}
		{\begin{tabular}{llllll}
		\hline
		\hline
		\ &    $\tilde{X}(0,8^{0},2)$ & $^2\Pi\,$(6819.3) & $^2\Sigma\,$(7087.6)& $^2\Pi\,$(7109.6) & $\tilde{X}(1,2^0,2)$ \\
		\   & $^{2}\Sigma ^{+, c}$ & $^2\Pi$\supd & $^{2}\Sigma ^{+, d}$ & $^{2}\Pi ^c$ & $^{2}\Sigma ^{+, d}$\\
		\hline
		$A$                     &           & -9.3640(85)  &                & -4.29(2)      &                     \\
		$B$                     & 1.42115(5)&1.37263(18)    & 1.3940939(21) & 1.3719(1)     & 1.430515(26)          \\
		$10^5\, D $      & 1.582(3)      & -0.31(20)\supe &  0.89473(97)    & 0.68(4)       & -7.850(16)           \\
		$10^2\, \gamma $ & 0.0015(2)   & 3.03(11) & -0.73636(67)   & -0.0336(6)    & -0.1940(81)           \\
		$10^2\,p+2q$                    &      & -3.47(20)  &               & -9.25(8)    &                      \\
		$10^3\,q$                       &      &2.89(15)       &               & -4.47(3)   &                       \\
		Origin                  & 6695.678(2)&6819.2687(41) & 7087.64950(8) & 7109.646(6) & 7527.0986(9)$^f$        \\
		\hline
		&\multicolumn{3}{c}{Calculated values$ ^{g} $} & \\
		\hline
		Origin  & 6696.7 & 6824.8 & 7099.0  & 7106.7 & 7547.7 \\		
		\hline
		\hline
		\multicolumn{6}{l}{\textit{a.} State labels from Ref.\cite{Tarroni2003} are: $\tilde{X}(2,1^{1},0)\tilde{A}(0,0,2)^{1}$,}\\
		\multicolumn{6}{l}{$\tilde{X}(0,3^{1},3)\tilde{A}(0,0,2)^{1}$, and $\tilde{X}(0,3^{1},3)\tilde{A}(0,0,2)^{1}$ for levels at 6819.3,}\\
		\multicolumn{6}{l}{ 7087.6 and 7109.6 \wn, respectively. The numbers in parenthesis are one }\\ 
		\multicolumn{6}{l}{standard deviation of the fit in units of the last quoted decimal place.}\\
         \multicolumn{6}{l}{\textit{b.} Lower state parameters are given in Table \ref{table1}.}\\
        \multicolumn{6}{l}{\textit{c.} Taken from Ref. \cite{Le2016}}\\
        \multicolumn{6}{l}{\textit{d.} Improved or newly determined parameters.}\\
        \multicolumn{6}{l}{\textit{e.} Higher order parameters determined for this state: $H=1.46(55)\times 10^{-8}$,} \\
        \multicolumn{6}{l}{$(p+2q)_D=-1.138(22)\times 10^{-3}$ and $q_D=1.10(13)\times 10^{-5}$.}\\
        \multicolumn{6}{l}{\textit{f.} Error does not include error in lower state energy in Table \ref{table1}.}\\
        \multicolumn{6}{l}{\textit{g.} Predicted values are from Tarroni and Carter.\cite{Tarroni2003}} \\	
	\end{tabular}}
	\normalsize
\end{table}


\clearpage
\newpage
\setcounter{figure}{0}
\renewcommand{\thefigure}{S\arabic{figure}}

\end{document}